\documentclass[twocolumn,superscriptaddress,10pt,aps,showpacs,reprint]{revtex4-1}

\usepackage{amsmath}
\usepackage{amssymb}
\usepackage{graphicx}
\usepackage{url}
\usepackage{longtable}

\begin{document}

\title{Collective behavior and emergent risks in a model of human- and autonomously-driven vehicles}

\author{Skanda Vivek}
\email{skanda.vivek@physics.gatech.edu }
\affiliation{School of Physics, Georgia Institute of Technology, Atlanta, GA 30332}
                                                        
\author{David Yanni}
\email{dyanni3@gatech.edu}
\affiliation{School of Physics, Georgia Institute of Technology, Atlanta, GA 30332}

\author{Peter J. Yunker}
\email{peter.yunker@gatech.edu}
\affiliation{School of Physics, Georgia Institute of Technology, Atlanta, GA 30332}

\author{Jesse L. Silverberg}
\email{jesse.silverberg@gmail.com}
\affiliation{Wyss Institute for Biologically Inspired Engineering, Harvard University, Boston, MA, 02115}
\affiliation{Department of Systems Biology, Harvard Medical School, Boston, MA 02115}

\begin{abstract}
While much effort has been invested in studies of traffic flow as a physics problem, two emerging trends in technology have broadened the subject for new investigations.  The first trend is the development of self-driving vehicles.  This highly-anticipated shift from human- to autonomous-drivers is expected to offer substantial benefits for traffic throughput by streamlining large-scale collective behavior.  The second trend is the widespread hacking of Internet-connected devices, which as of 2015, includes vehicles.  While the first proof-of-concept automobile hack was done at the single-vehicle scale, undesirable collective effects can easily arise if this activity becomes more common.  Motivated by these two trends, we explore the phenomena that arise in an active matter model with lanes and lane-changing behavior.  Our model incorporates a simplified minimal description of essential differences between human- and autonomous-drivers.  We study the emergent collective behavior as the population of vehicles shifts from all-human to all-autonomous.  Within the context of our model, we explore a worst-case scenario where Internet-connected autonomous vehicles are disabled simultaneously and \textit{en masse}.  Our approach reveals a model-independent role for percolation in interpreting the results.  A broad lesson our work highlights is that seemingly minor malicious activity can ultimately have major impacts when magnified through the action of collective behavior.
\end{abstract}

\maketitle
Autonomous vehicles are highly sought for their potential to revolutionize transportation.  This convergence between automotive and IT industries is expected to ripple beyond the immediate in-vehicle experience, and yield a broader array of benefits for society with changes in both commercial and private sectors\cite{litman2014} [see SM].  However, this focus on the value of technology-driven disruption leaves potential blind spots with respect to emergent risks brought by automotive computerization.  For example, proof-of-concept hacks in 2015 and 2016 demonstrated the ability to remotely control an Internet-connected vehicle, including the brakes, acceleration, and steering \cite{Greenberg2015Hackers, Greenberg2016The}.  Given that the level of computerization will only increase with self-driving vehicles, there is a deep need to understand the consequences of hackers leveraging remote automation to achieve undesirable ends.  We therefore consider a worst-case scenario where various numbers of connected vehicles are simultaneously and \textit{en masse} disabled during transit.  While the details of how such a hack would be executed are beyond the scope of our work here, similar concerns have been raised by cybersecurity experts\cite{parkinson2017cyber, amoozadeh2015security}. To explore this critical blind spot, we model human and autonomously driven vehicles as an active matter system\cite{ramaswamy2010mechanics, marchetti2013hydrodynamics}.  Unlike conventional flowing systems subject to the kinetic phenomenon of clogging\cite{thomas2015fraction, zuriguel2014clogging}, the incorporation of lanes and lane changing behavior motivates the geometric phenomenon of percolation as the relevant framework for understanding collective behavior that emerges in the presence of disabled vehicles.

We begin this study by defining a microscopic model for human and autonomously driven vehicles.  Of the different approaches to modeling vehicular collective motion\cite{helbing2001, treiber2013, lwr, newell2002, treiber2000, nagel1992, nagatani2002physics,ovm-bando1994}, we chose a force-based approach\cite{silverberg2013collective, bottinelli2016emergent, bottinelli2017using} that represents each vehicle by their position $x$ at time $t$ on a straight road of length $L$.  The main appeal of this type of model is the rich complexity of interesting collective behavior that can be generated by its simple set of equations and correspondingly small set of parameters.  Numerical simulations were performed using velocity Verlet integration for $N$ vehicles seeded with uniformly distributed random initial positions, zero initial velocity, zero initial acceleration, and periodic boundary conditions.  The position of each vehicle was evolved using a self-propulsion force $F^{\rm propulsion}$, and a repulsive collision-avoidance force $F^{\rm repulsion}$ according to
\begin{eqnarray}
\ddot{x} & = & F^{\rm propulsion} + F^{\rm repulsion}, \nonumber \\
F^{\rm propulsion} & = & \tau_{\alpha}^{-1} (v_{\alpha} - \dot{x}), \nonumber \\
F^{\rm repulsion} & = &  \left\{ 
\begin{array}{ll}
\epsilon_{\alpha} \left( 1 - \delta x/R \right)^{3/2}, & \delta x < R \\
0, & {\rm otherwise}
\end{array}	\right. 
\label{eq1}
\end{eqnarray}
where $\dot{x}$ and $\ddot{x}$ are the velocity and acceleration of a given vehicle [see SM].  Each vehicle has a preferred speed $v_{\alpha}$ and a characteristic response time $\tau_{\alpha}$ that it takes to equilibrate to this speed.  In the limit of a single vehicle without interactions, $F^{\rm repulsion} \equiv 0$ and direct integration shows $\dot{x}(t) = v_{\alpha} (1 - e^{-t/\tau_{\alpha}}).$  As the number of vehicles increases and they begin to interact, the repulsive force slows a given vehicle down as it approaches another vehicle from behind.  The strength of this force increases smoothly from zero as the vehicle-vehicle separation $\delta x$ becomes smaller than the interaction threshold distance $R = (29~{\rm m/s}) \cdot(2~{\rm s})$.  This threshold models the 2-second rule on highway-like conditions, which states a minimum safe distance between two vehicles is the distance traveled during 2 seconds at typical speeds of 29~m/s (65 mi/h).  The functional form of the repulsion force in Eqs.~(\ref{eq1}) follows from the compression of two elastic spheres\cite{landau1959course}, and only applies to the trailing vehicle, as drivers respond to traffic ahead of them more strongly than behind.  Under these conditions, we balance propulsion and repulsion forces when two vehicles are bumper-to-bumper, and set $\delta x = r \equiv 4.5$ m as the typical size of a vehicle to find $\epsilon_{\alpha} \equiv [ (29~{\rm m/s}) / \tau_{\alpha}] ( 1- r/R )^{-3/2}.$  Thus, the distance between vehicles $\delta x$ is always larger than the size of a vehicle $r$, but the interaction between vehicles is only experienced when $r \le \delta x < R$.  

For simplicity, we focus on the emergent collective motion of two types of vehicles: human-driven and fully autonomous.  In Eqs.~(\ref{eq1}), this simplification amounts to distinct values for the variables with subscript $\alpha$, where we use $\alpha = H$ for human-driven vehicles and $\alpha = A$ for autonomous vehicles.  For example, $N_H$ denotes the number of human drivers and $N_A$ denotes the number of autonomous drivers.  Each of the $N_H$ human drivers is given a randomly chosen preferred velocity $v_H$ chosen from a Gaussian distribution with mean 29 m/s and standard deviation 6.7 m/s (15 mi/h).  For the $N_A$ autonomous drivers, we fix their preferred velocity $v_A = 29$ m/s to be constant and equal to the typical speed limit.  This homogeneity reflects the notion that autonomous vehicles will be engineered to travel at the legally allowed maximum speed to optimize traffic flow and safety.  

\begin{figure}
\includegraphics[scale=.98]{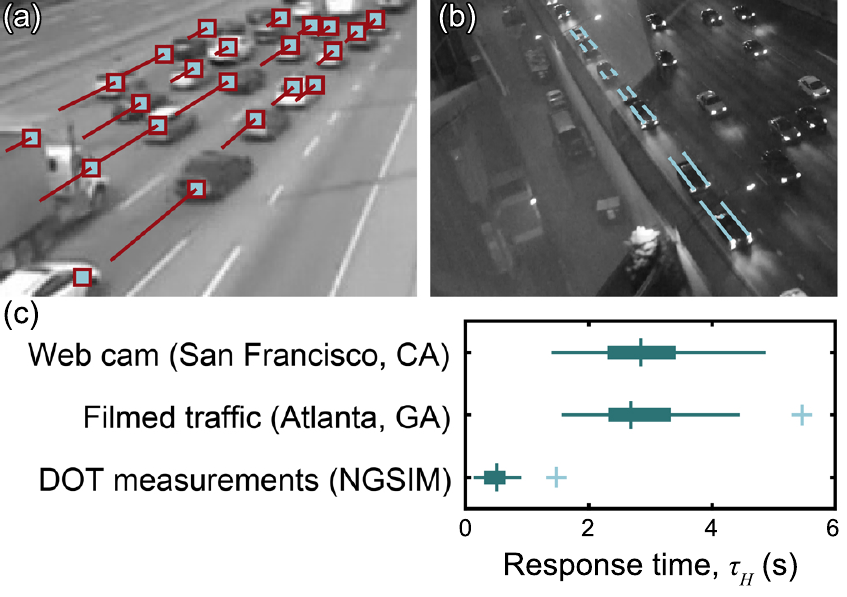}
\caption{Empirical measurements of a human-driven vehicle's response time $\tau_H$.  (a) A live streamed web cam provides daytime video data on traffic flow (northbound 101 at N 1$^{\rm st}$ street, San Francisco/Oakland Bay, CA).  Individual vehicles (squares) and their trajectories (lines) illustrate measurements of $x(t)$.  (b) Nighttime highway traffic was filmed and analyzed using the same methods (southbound I85 exit 249C, Atlanta, GA).  (c) Fitting velocities during stop-and-go motion allows for measurements of the human-driver response time $\tau_H$.  Box-and-whisker plots show the median value, data quartiles, and outliers (+). }
\label{fig1}
\end{figure}

The only remaining free parameter in the equations of motion for human-driven vehicles is the response time $\tau_H$, which generally varies from 0.5~s to several seconds, depending on attentiveness, driving conditions, visibility, and level of distractions. To empirically measure $\tau_H$, we analyzed three sources of driving data including a highway webcam [Fig.~\ref{fig1}(a)]\cite{livecam}, digital footage we recorded [Fig.~\ref{fig1}(b)], and previously published Department of Transportation (DOT) traffic data from Los Angeles, California (southbound US 101; NGSIM June 15, 2005, 7:50 to 8:05 am)\cite{ngsim2005, alexiadis2007model, montanino2013making}.  By measuring the velocity in ``stop-and-go'' traffic and assuming the repulsion force is negligible during the initial ``go,'' we could fit this data to $\dot{x}(t)/\dot{x}(t_f) = 1 - e^{-t/\tau_H}$ resulting in measurements of $\tau_H$ [Fig.~\ref{fig1}(c)].  In higher-density traffic, we observed drivers tend to accelerate over longer periods of time, whereas in lower-density traffic, drivers tend to react faster.  Because Eqs.~(\ref{eq1}) simplifies driver response time to be density-independent, we set $\tau_H = 2.0$~s, which comes from the average of empirical measurements that were found to be $(2.0 \pm 1.4)$~s.

\begin{figure}
\includegraphics[scale=.98]{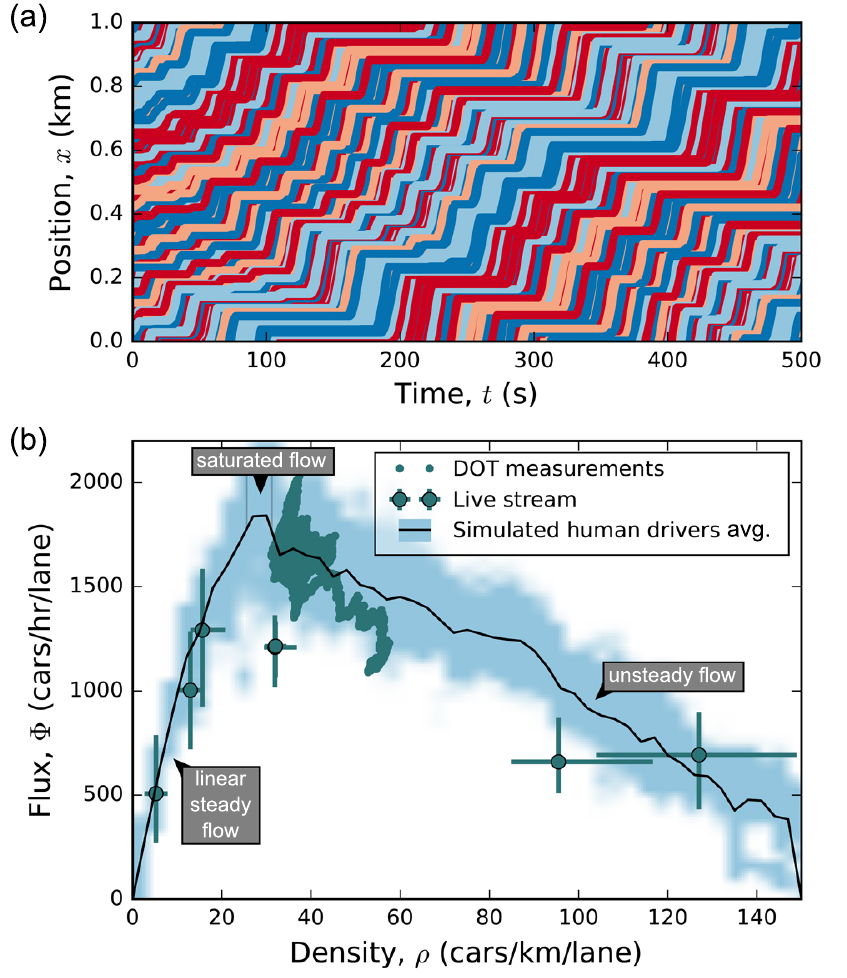}
\caption{Simulations of human-driven vehicles using Eq.~(\ref{eq1}) reproduce empirical data.  (a) Trajectory plots of $N_H = 100$ simulated human drivers on a $L = 1$ km, $\ell = 1$ lane road.  Each colored line corresponds to a vehicle that generally increases its position $x$ over time.  Horizontal line segments indicate stand-still conditions with no forward motion.  Bands of horizontal line segments that move diagonally-down on the plot are phantom traffic jams, which manifest as backwards-propagating density waves.  (b) Flux-density relationship for human-driven vehicles on a $L = 1$ km, $\ell = 3$ lane road shows favorable agreement between empirical and simulation results.  Error bars on live stream data are the inter-quartile range.  DOT measurements are plotted as a continuous series of points with density $\rho$ greater than the critical density $\rho^* \approx 30$ cars/km/lane.  Solid line is the simulated mean value of $\Phi(\rho)$ and the shaded band shows statistical fluctuations.}
\label{fig2}
\end{figure}

In simulations with $N_H = 100$ human-driven vehicles on an $\ell = 1$ lane road, $L = 1.0$ km long, we find a brief transient regime that lasts for $\approx 500$ integration time steps, followed by steady-state dynamics.  Superimposed on this 1D line of traffic, we observe backwards-propagating density waves known as ``phantom traffic jams'' that cause stop-and-go motion of individual vehicles [Fig.~\ref{fig2}a].  This emergent collective phenomenon has been extensively studied in other models of vehicle traffic\cite{helbing2001, treiber2013, lwr, newell2002, treiber2000, nagel1992, nagatani2002physics}, and is generally known to arise from the non-zero driver response time $\tau_H$.  Here, we see these density waves as an important validation of our active matter model, as their spontaneous emergence indicates broad consistency with the known phenomenology of vehicle traffic.

Going beyond one-lane simulations and reproducing essential features of multi-lane traffic where $\ell > 1$ requires a model for lane changing behavior.  Otherwise, quantitative metrics of traffic flow would be dominated by the slowest moving vehicle.  One method for realistically capturing these dynamics is the Minimizing Overall Breaking Induced By Lane changes (MOBIL) framework\cite{treiber2006mobil, treiber2009modeling}, which offers a simple force-based rule to determine when a vehicle should change lanes.  Specifically, MOBIL considers whether a vehicle and its neighbors would better match their preferred speeds if a given vehicle changes lanes [see SM].  In our implementation, this algorithm effectively minimizes $F^{\rm repulsion}$ over all vehicles involved, and enables quantitative predictions of traffic flow beyond the qualitative observations of phantom traffic jams [Fig.~\ref{fig2}(a)].  For example, we use the density-dependent flux $\Phi(\rho)$ as a measure to study the emergent collective properties of multi-lane traffic flow [Fig.~\ref{fig2}(b)].  In empirical measurements and simulations with $\ell = 3$ lanes, we use a portion of the road $l \approx 500$~m to calculate $\Phi = \sum_i \dot{x}_i / (\ell l)$ and $\rho = \sum_i 1 / (\ell l)$, where the index $i$ runs over all vehicles in $l$, sampled at statistically independent temporal intervals.  Comparing observational data [Fig.~\ref{fig2}(b), green dots] to simulation predictions with MOBIL lane changing [Fig.~\ref{fig2}(b), black line and blue-shaded band] shows a broad consistency with no fitting parameters.  Both in empirical measurements and simulations, we find that when the density is low, $\Phi \propto \rho$ [Fig.~\ref{fig2}(b), $\rho < 30$ cars/km/lane].  As the density increases to a critical value $\rho^* \approx 30$ cars/km/lane, the flux peaks, and subsequently declines for increasing $\rho$.  In simulations, we find $\rho^*$ corresponds to an average of one vehicle per 33~m of road per lane.  This distance is $\approx 1.1$~s of travel time between vehicles, placing them within the repulsive-force interaction distance $R$.  As such, when $\rho > \rho^*$, the system is effectively a 1D continuum with density waves [Fig.~\ref{fig2}(a), horizontal bands] that produce the decline of $\Phi$ at larger $\rho$.

\begin{figure}
\includegraphics[scale=.98]{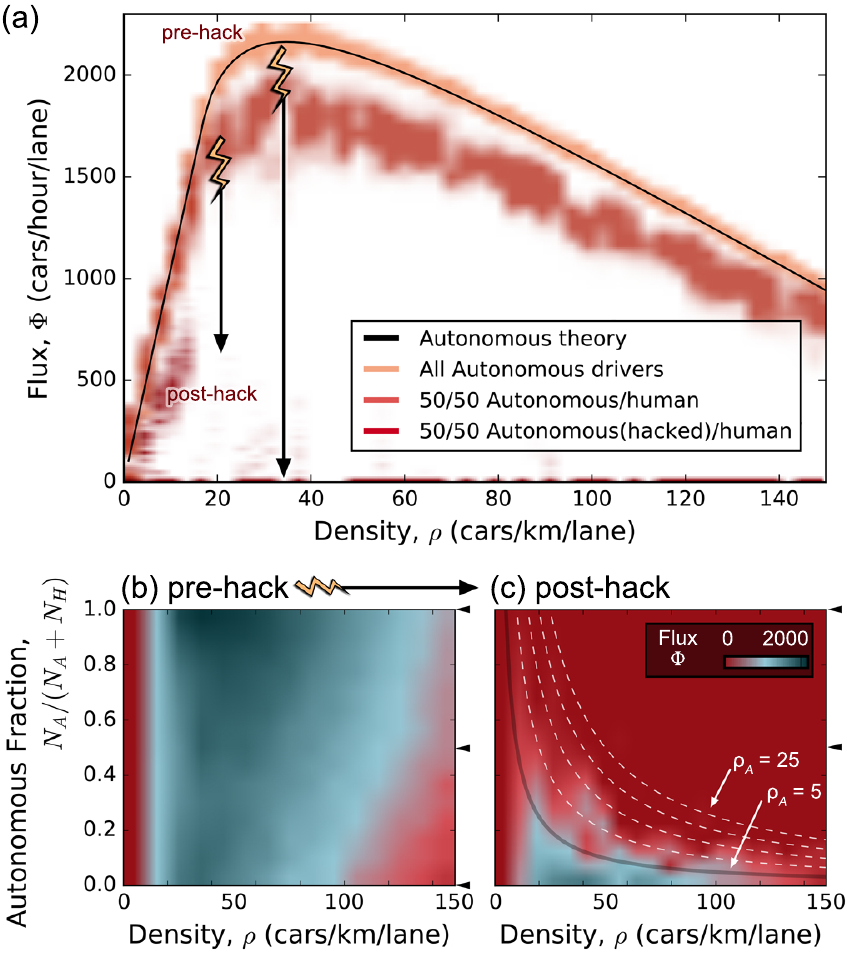}
\caption{While increasing the number of autonomously-driven vehicles generally increases the total flux $\Phi$, a malicious hack disabling autonomous vehicles sends $\Phi$ to near-zero values, especially in high-density traffic.  (a) Flux-density relationship shows how increasing the number of autonomous drivers from 50\% (upper light-red data) to 100\% (upper light-bronze data) offers greater vehicle throughput.  A hack that disables autonomous vehicles reduces the flux to zero or near-zero values (lower dark-red data).  Solid black line is a theoretical prediction for autonomous drivers based on force-balance in Eq.~\ref{eq1}.  (b) Heat map for the flux $\Phi$ generally shows increased throughput of vehicles as autonomous drivers replace human drivers.  (c) Disabled autonomous vehicles create bottlenecks on the road leading to unsteady and near-zero flow.  Gray lines represent contours of constant density of disabled autonomous vehicles $\rho_A = 5, 10, 15, 20,$ and $25$.}
\label{fig3}
\end{figure}

Thus far, our simulations and analysis have focused on human-driven vehicles.  Incorporating autonomous vehicles requires that we fix the final parameter $\tau_A$, which is the autonomous driver response time.  Given these highly computerized vehicles are expected to utilize a combination of local sensor information, wirelessly shared non-local traffic conditions, and cloud-connected AI, we can reasonably assume they will respond to driving conditions more rapidly than human drivers.  As an order of magnitude estimate, we set the autonomous vehicle response time $\tau_A = 0.1$~s, which is a fraction of the human response time $\tau_H$, but still larger than the simulation time step.  Simulations with all autonomous vehicles immediately show important differences in the flux $\Phi(\rho)$, highlighting a key advantage of autonomous vehicles over human drivers that underlies their anticipated impact on automotive transport.  Specifically, $\Phi(\rho)$ still peaks near the same critical density $\rho^*$, however, with all autonomous vehicles $\Phi(\rho^*)$ is nearly 20\% larger [Fig.~\ref{fig3}(a), upper light-bronze data].  Even with a 50/50 mix of autonomous and human vehicles there are substantial advantages at the highest densities where the flow is steadier relative to results with all human drivers [Fig.~\ref{fig3}(a), upper light-red data, $\rho > \rho^*$ cars/km/lane].  The combination of higher throughput and steadier flow translate to reduced congestion, reduced fuel consumption, and less time spent traveling between destinations.  

Interestingly, $\Phi(\rho)$ in all-autonomous vehicle simulations can be predicted from Eqs.~(\ref{eq1}).  Given the uniform preferred speed and rapid response time, $\Phi \approx N_A \dot{x} / (\ell L) = \rho \dot{x}$.  Under steady conditions, propulsion and repulsion forces balance so that $\ddot{x} = 0$ and $\delta x = 1/\rho$, which leads to $\dot{x} = v_A - \tau \epsilon_A [1 - 1/(2 \rho v_A)]^{3/2}$.  The flux is therefore the real part of $\Phi(\rho) = \rho v_A - \rho \tau \epsilon_A [1 - 1/(2 \rho v_A)]^{3/2} = \rho v_A \left[ 1 - \left( \frac{1 - 1/(2\rho v_A)}{1 - r/R} \right)^{3/2} \right]$, where we have used the expression for $\epsilon_A$ derived from force balance to find the final equation.  Comparing this prediction to simulations shows strong agreement [Fig.~\ref{fig3}(a), solid black line].

From the perspective of the flux-density relationship, there are measurable benefits to having an increased number of autonomous vehicles on the road due to the generally increased throughput of traffic.  However, as indicated by recent proof-of-concept hacks, Internet-connected vehicles are endowed with enhanced abilities at the cost of new risks.  The worst-case scenario we consider is one in which hacked vehicles are disabled during transit, which in the context of our model, amounts to setting $v_A = 0$ after the simulation has reached steady-state.  This type of event has actually been proposed as an active self-defense mechanism for vehicles that detect their systems have been compromised, though it was offered in the context where only a single vehicle was being considered\cite{parkinson2017cyber}.  If we assume this strategy was broadly deployed, a wide-spread hack that compromises many vehicles simultaneously would most certainly give rise to new collective behavior that affects any functional vehicles still on the road.  For example, a 50/50 mixture of human and autonomous drivers in this circumstance has a substantial loss of flow with a nearly 60\% reduction in the peak flux and a total loss of flow at higher densities [Fig.~\ref{fig3}(a), lower dark-red data $\rho > 20$ cars/km/lane].  

Generating heat maps for $\Phi(\rho)$ with varying fractions of autonomously driven vehicles $N_A / (N_A + N_H)$ offers a more complete view of the benefits of autonomous vehicles [Fig.~\ref{fig3}(b)] as well as the effects on flow when these vehicles are disabled [Fig.~\ref{fig3}(c)].  Comparing the pre-hack to the post-hack results show the flow largely collapses when vehicles are disabled, leaving behind unsteady and near-zero flux, even at relatively low fractions of autonomous vehicles [Fig.~\ref{fig3}(b) and (c), increased size of red region].  In the post-hack scenario we observe the flux substantially drops when the fraction of disabled autonomous vehicles exceeds 10 to 20\%  [Fig.~\ref{fig3}(c), boundary of lower green band].  At the same time, this number of self-driving vehicles corresponds to a rather small benefit in the pre-hack scenario relative to bounds at the $N_A / (N_A + N_H) = 0$ and $1$ limits [Fig.~\ref{fig3}(b)].  Evidently, the risks introduced by connected-car technologies have an earlier onset than the benefits, as measured by vehicle throughput.

\begin{figure}
\includegraphics[scale=.98]{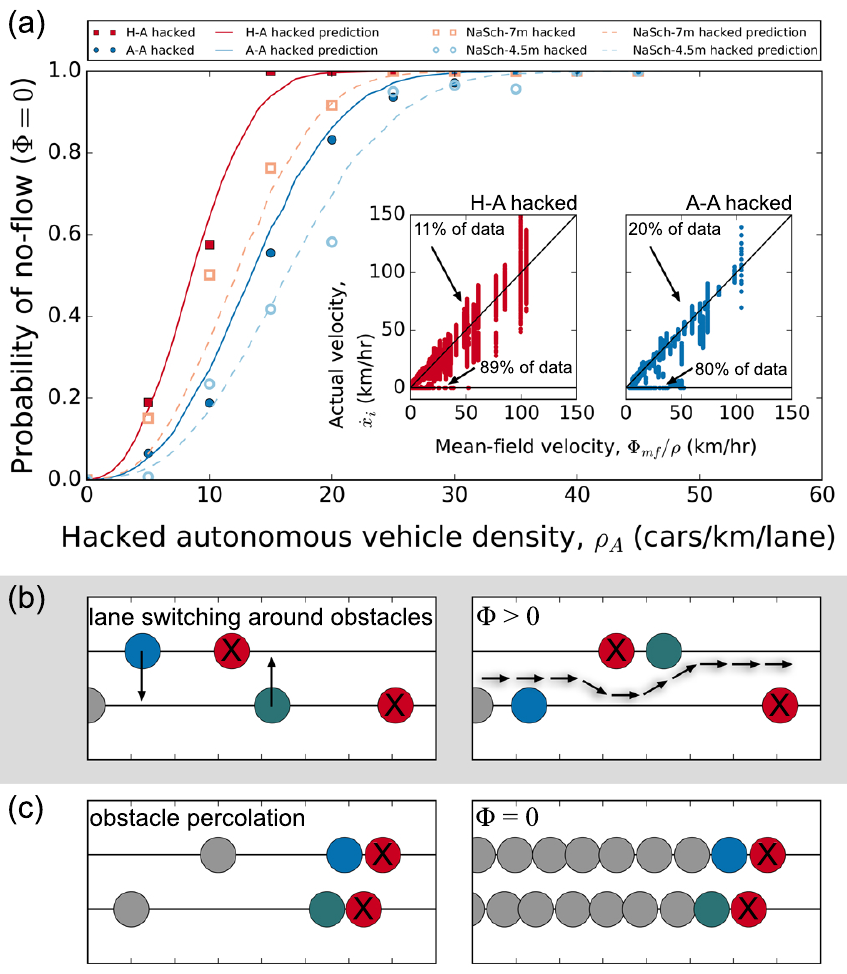}
\caption{Percolation of obstacles is a generic mechanism for loss of flux.  (a) Probability of no-flow conditions ($\Phi = 0$) at a given density of disabled autonomous vehicles.  Simulation data for combinations of human-driven and hacked-autonomous vehicles (H-A, solid red squares) are consistent with a geometric obstacle percolation prediction (H-A, solid red line).  Similarly, simulation data for uncompromised and hacked-autonomous vehicles (A-A, solid blue circles) are consistent with the same geometric percolation prediction (A-A, solid blue line).  These findings are reproduced by the Nagel-Schrekenberg cellular automata model with a population of disabled vehicles (red and blue open symbols), and likewise agree with the predictions of geometric obstacle percolation (red and blue dashed lines).  Left inset shows the velocity of each human-driven vehicle in H-A simulations versus the mean-field predicted velocity.  The data separates onto two lines indicating two discrete outcomes with $\Phi > 0$ (black diagonal line, slope = 1), and $\Phi = 0$ (black horizontal line, slope = 0).  Right inset shows the same plot for simulations of all-autonomous vehicles, where only a fraction have been disabled.  Here again, we find the same set of discrete outcomes.  Cartoon illustrations of model showing (b) how lane switching around non-percolated obstacles allows for flux $\Phi > 0$, and (c) how percolated obstacles lead to a complete loss of vehicle flow ($\Phi = 0$).}
\label{fig4}
\end{figure}

To better understand the emergent collective behavior of the scenario being considered, we note the resulting flux contours of the post-hack heat map appear to follow lines of constant autonomous vehicle density $\rho_A$ [Fig.~\ref{fig3}(c), solid and dashed lines].  This observation suggests $\rho_A$ is a quantity of significant explanatory power for how disabled vehicles affect traffic flow.  We therefore use the data that generated the heat map to calculate the probability that $\Phi = 0$ as a function of $\rho_A$, revealing a well-defined sigmoidal-like relationship [Fig.~\ref{fig4}(a), solid red squares].  Furthermore, when we plot the actual velocity of vehicles $\dot{x}_i$ against the mean-field velocity $\Phi / \rho$, we see the data separates onto two distinct curves: the $\dot{x}_i = \Phi / \rho$ curve where simulation and prediction agree up-to fluctuations [Fig.~\ref{fig4}(a) left inset, red data along black diagonal line, 11\% of simulation data], and the $\dot{x}_i = 0$ no-flow curve [Fig.~\ref{fig4}(a) left inset, red data along black horizontal line, 89\% of the simulation data].  Taken together, the combination of the bimodal separation of velocities and the smooth sigmoidal relationship between flow and no-flow outcomes strongly suggests the presence of a percolation phase transition.  

A simple description of percolation is the study of flow through porous materials\cite{kirkpatrick1973percolation, aharony2003introduction, bollobas2006percolation, silverberg2009model}.  When the material's pores are connected such that a continuous path exists between two surfaces, fluid is able to move through the material.  Conversely, when no continuous path between surfaces can be found, fluid cannot flow.  Complex phenomenology is then introduced through the non-trivial network of connections among the material's pores.  Often, random lattice models are used to recapitulate the statistical properties of pore connectivity, and the prominent features studied are the emergence of a flow to no-flow phase transition where universality and scaling phenomena are readily found.  

In our model, there appears to be a percolation phase transition where the flow of traffic is analogous to the flow of fluid, the $\ell$-lane road is the material, and disabled vehicles create a network of obstacles reminiscent of a series of connected pores.  Freely-moving vehicles approaching a disabled vehicle from behind continue to flow, as would a conventional fluid, although the underlying mechanism relies on lane switching behavior [Fig.~\ref{fig4}(b)].  When disabled vehicles align across all $\ell$ lanes, they form an immobile barrier obstructing flow [Fig.~\ref{fig4}(c)].  In this sense, the percolation of obstacles appears to be the critical factor determining traffic flow properties, especially when asking whether a given simulation produces either a flow or no-flow condition [Fig.~\ref{fig4}(a), left inset, diagonal and horizontal branches].  Obstacle percolation should be differentiated from the formation of clogs\cite{nagatani2002physics, cejas2017particle}, which are dynamic in nature, depend on observation time, and have statistical properties distinct from percolation that do not match our results.  Nevertheless, if percolation is the correct framework, then stochastically placing obstacles at density $\rho_A$ on an $\ell$-laned road and computing the probability that these obstacles percolate should reproduce the measured probability distribution for $\Phi = 0$.  In this calculation, we set the length for steric interactions of disabled autonomous vehicles and moving human-driven vehicles equal to the balance of repulsion and propulsion forces, described by Eqs.~(\ref{eq1}).  On average, these values are 8.5~m ($> r$) for human-driven vehicles and 4.5~m ($ = r$) for autonomous vehicles, where the larger-than-expected average value for human-driven vehicles arises from heterogeneity in the preferred velocity $v_H$ and the non-linear repulsion force.  Remarkably, this minimal calculation, which only depends on the geometry of percolation, is able to accurately reproduce the probability distribution for $\Phi = 0$ [Fig.~\ref{fig4}(a), solid red line].

In light of the different effective sizes for human and autonomous vehicles, we can use geometric obstacle percolation to predict the probability that $\Phi = 0$ when the road contains all autonomous vehicles, but only a fraction are disabled [Fig.~\ref{fig4}(a), solid blue line].  Comparing this new scenario to a direct microscopic simulation again shows a bimodal separation of flow and no-flow outcomes [Fig.~\ref{fig4}(a), two branches of right inset].  Moreover, the probability distribution for $\Phi = 0$ computed separately by microscopic simulation and geometric obstacle percolation accurately agree with one another [Fig.~\ref{fig4}(a) solid blue circle match solid blue line].  We can further show the emergence of percolation is a model-independent result when explaining the effects of disabled vehicles.  We implemented the Nagel-Schrekenberg cellular automata model of traffic flow on an $\ell$-lane road with steric interaction of 4.5 and 7~m, disabled a number vehicles at density $\rho_A$ by setting their speed to 0, and computed the probability that $\Phi = 0$.  Even in this completely different model for traffic flow, the microscopic numerical simulations [Fig.~\ref{fig4}(a), open red squares and circles] continue to match the prediction of obstacle percolation [Fig.~\ref{fig4}(a), dashed red and blue lines].  Thus, as as suggested by the contours of constant $\rho_A$ in the post-hack heat map for $\Phi$ [Fig.~\ref{fig3}(c)], the density of autonomous vehicles $\rho_A$ explains the emergent flow phenomenology.

We motivated this work by exploring the unintentional collective effects that can arise when Internet-connected autonomous vehicles are disabled suddenly and \textit{en masse}.  Our approach utilized a lane-based active matter model and uncovered a model-independent role for percolation in interpreting the results.  Looking more broadly at the immediate consequences of obstacle percolation, we see a striking set of potential outcomes: emergency vehicles would be unable to respond to calls for help, food shipments to grocery stores would be delayed, and long-distance commuters would be unable to get to work.  Urban cores and suburban towns, which are key centers of economic activity, would cease to function normally.  Given the magnitude of these potential harms, there is a clear need to proactively develop strategies that mitigate the risks posed by obstacle percolation.  

Fortunately, our results on the statistics of obstacle percolation indirectly suggest a way to achieve this risk-mitigation.  The probability of no-flow is small when the densities of disabled vehicles are correspondingly low [Fig.~\ref{fig4}(a)], so the goal should be to minimize the number of affected vehicles, as one would naturally guess.  The probability of geometric percolation allows us to assess how different strategies result in different effective risk levels. One approach could be to distribute vehicle communications across multiple independent networks in such a fashion that a hack affecting one network would have no effect on the others.  These multiple non-overlapping networks would appear inefficient from the perspective of communication standards, but it offers a greater degree of robustness with respect to the cyber-physical issues motivating this study.  While such strategies are yet to be developed and deployed, only by anticipating these worst-case scenarios can we inoculate for their consequences.  For example, if there were 10 separate networks, each connected to 1/10 of all vehicles, then a network-wide hack would have the effect of disabling only 10\% of the vehicles on the road.  In light of our simulation results [Fig.~\ref{fig3}(c), $N_A / (N_A + N_H) = 0.1$], this relatively modest number of disabled vehicles would still allow traffic to generally flow.  To achieve higher impact, a hacker would have to breach multiple separate networks, each with their own security and communications protocols, increasing the level of difficulty and sophistication required to realize a worst-case scenario.  While these particular simulations were computed with specific ratios of human and autonomous drivers, the physics of percolation [Fig.~\ref{fig4}(a)] ensures similar outcomes.  Indeed, this generality is the strength of the obstacle percolation interpretation uncovered here.

\section*{Acknowledgments}
The authors thank Zohar Nussinov for useful discussions and Christopher Giottonini for assistance with video acquisition.  SV, DY, and PJY acknowledge support from the Georgia Tech Soft Matter Incubator.  JLS was independently funded.

%
\setcounter{equation}{0}
\setcounter{figure}{0}
\renewcommand{\theequation}{S\arabic{equation}}
\renewcommand{\thefigure}{S\arabic{figure}}
\renewcommand{\thetable}{S\arabic{table}}

\begin{widetext}
\section*{Supplemental Materials}
\end{widetext}

\section*{Foreseeable ripple effects of autonomous vehicles}
With any major technological innovation, there are foreseen and unforeseen consequences that ripple through society.  To date, considerable effort has gone into studying how the U.S. public and private sectors will be affected by autonomous vehicles\cite{recode2017auto, cnbc2017self, wapo2016how, center2016auto, npr2015map}.  Here, we briefly summarize these reports to provide further context for the motivation of our work and to more broadly indicate the potential consequences of a malicious hack on autonomous vehicles.
\begin{enumerate}
\item \textbf{Transportation as a Service (TaaS):}  Individual ownership of vehicles is a standard model for personal transportation, especially away from urban centers where public transit systems are generally unavailable.  Autonomous vehicles are anticipated to shift away from the ownership model where transportation is a ``product'' and toward transportation as an on-demand service\cite{recode2017auto}.  Generally, the idea of TaaS is that individual riders do not own the vehicle they ride in.  Instead, they pay a fee to be transported between two destinations.  While this change in ownership model is expected to bring travel costs down, it also means the vehicle can be immediately used by another individual in need of transportation services.  Numerous automotive manufacturers are already beginning to move their business models in this direction with pilot programs targeting the ride-sharing market.  
\item \textbf{Parking:}  Following from the notion of TaaS, parking requirements for individually owned vehicles will wane.  One estimate concluded that approximately 60 of the 150 billion ft$^2$ of total parking (including parking garages, lots, street parking, etc.) in the U.S. will be available for redevelopment\cite{cnbc2017self}.  However, this shift in land use also means that municipal revenues from parking ticket fines will also be reduced.  For large cities, such as Los Angeles and New York, this loss is estimated around $\$150$ and $\$550$ million.  
\item \textbf{Law Enforcement:} Loss of revenue for law enforcement services due to the reduction in parking tickets is expected to be accompanied by decreased revenues from speeding and other traffic violations.  As a recent example from 2014, Washington D.C. issued an average of 773 tickets per day from speeding cameras alone, totaling $\approx \$40$ million per year\cite{wapo2016how}.  In more extreme cases, a few small cities and municipalities in the U.S. generate nearly 50\% of the town revenue through traffic tickets\cite{wapo2016how}.  It remains an open question how these budgetary losses will be dealt with.  Along similar lines, there is a reasonable expectation that the focus for police activity will also shit; traffic officers and emergency responders will see decreased demand, leading to either job losses or reallocation of police staff.    
\item \textbf{Insurance:} One of the most anticipated benefits of autonomous vehicles is a decrease in the number of automotive accidents.  While this shift reduces demand for emergency responders, it will also significantly rewrite the automotive insurance industry's financial model.  Instead of a monthly fee based on driving history, we can reasonably expect to see more usage-based policies that charge by the number of miles traveled per year.  Similarly, health insurance policies will need to absorb changes in the number of automotive-related injuries.  For example, $\approx 37,000$ Americans die in car accidents each year, with nearly 100-fold more treated in hospital emergency rooms\cite{recode2017auto}.  This annual expenditure of $\approx \$30$ billion is likely to see dramatic reductions, directly affecting both the insurance industry and how hospitals fund their services.
\item \textbf{Energy Consumption:} A variety of factors that contribute to the total energy consumption of automotive vehicles are expected to change\cite{recode2017auto, center2016auto}.  Decreased congestion, eco-friendly driving algorithms that prioritize efficiency over performance, platooning, improved crash avoidance, and vehicle right-sizing are all expected to reduce the overall amount of energy consumed in autonomous vehicles.   Higher highway travel speeds, reduced travel costs, and increased access for poor, elderly, and disabled users, on the other hand, are expected to increase the amount of energy consumed.  Exactly how these two sets of competing forces balance remain to be seen, but a shift away from fossil fuels and towards electrification is broadly acknowledged.
\item \textbf{Real Estate:} Current land usage and valuation has been established with human-driven vehicles in mind.  However, if electric autonomous vehicles are intelligently programmed to return to a centralized charging and maintenance hub, then there will be a waning need for gas stations, consumer-facing vehicle maintenance shops, and car wash stations.  As these businesses see less demand, the property can be sold and repurposed for other needs.  Similarly, suburban and exurban communities will see shifts in property value as commutes become easier and faster.  
\item \textbf{Package and Food Delivery:} The U.S. Census Bureau has two broad categories in management and sales for employees not elsewhere classified.  Excluding these non-specific categories, trucking and delivery drivers employes 3.5 million people, making it the most common job in 27 of the 50 U.S. states\cite{npr2015map}.  As human drivers are replaced, the economic savings for delivery companies are expected to reduce consumer costs.  Comparisons with automation's effect in manufacturing suggest there will be a substantial mismatch in skills and geographic location between these unemployed drivers and available jobs.  As such, there will likely be a substantially increased demand for job retraining services.
\end{enumerate}

While further knock-on effects are expected to ripple out into additional sectors of public and private life, the examples provided here suggest a malicious hack that disables autonomously-driven vehicles would have broad social, economic, and security consequences.

\section*{A primer on malicious hacking}

A number of high-profile events illustrate the various ways hacking can be weaponized.  A selection of recent examples include: (i) distributed denial-of-service with zombie botnets, (ii) large-scale social engineering and disinformation campaigns using stolen credentials, (iii) remote deactivation of electricity grids\cite{Greenberg2017How}, (iv) industrial sabotage\cite{gjelten2010cyberworm, shapiro2016All}, (v) theft of money, identity, and intellectual property, (vi) disrupted access to medical facilities\cite{chappell2017ransomware}, (vii) disclosure of personal information as a means of coercion, and (viii) large-scale covert cyber-spying.  Individually, these malicious hacks demonstrate the harm a few individuals can inflict with the right software exploits.  Collectively, they form a broader picture of the social challenges presented by technology.  To further illustrate, we summarize some of the more common yet lower-profile forms of hacking prevalent today.
\begin{enumerate}
	\item Digital eavesdropping can be easily achieved with packet sniffers such as Wireshark or its equivalents.  These software tools can be used defensively to analyze and record suspicious network traffic, however, they can also be used to discretely monitor the activity of others without their consent.
	\item Keyloggers are a class of software that monitors user input through the mouse and keyboard.  Generally, the purpose is to discretely record sensitive information such as passwords.  
	\item Denial of Service (DoS) and Distributed Denial of Service (DDoS) attacks flood targeted servers in order to overload bandwidth and computational resources, forcing the target into a non-responsive state.  In DDoS attacks, hackers will deploy remotely controlled computers (bots) to increase the number of data packets used in the attack.  Interestingly, the damage inflicted by DDoS attacks are mediated through collective effects of multiple bots simultaneously flooding the target.  This example shows how collective properties have already been weaponized by malicious hackers.
	\item Brute force hacking is a resource-intensive attack typically launched against narrowly chosen targets.  It involves the systematic effort to access and decrypt data by trying all combinations of passwords.
	\item As an alternative to brute force, man-in-the-middle hacks can use various techniques to insert themselves between a user and the network.  From this position, an attacker can force or fool the user to enter sensitive data, passwords, or information of value.  
	\item Once installed, malware, viruses, and trojan software are routinely used in a variety of ways.  Common examples include (i) the covert transmission of information such as data from a keylogger, (ii) theft of intellectual property, and (iii) the installation of software that forces a system to discretely participate in DDoS attacks.
\end{enumerate}

In a relatively short period of time, hacking has gone from an esoteric pastime requiring in-depth expertise to an easily accessible outlet for malicious behavior.  Today, there are multiple front-end software packages enabling anyone with basic shell scripting knowledge to perform all of the attacks listed above.  Indeed, publicly available software such as Wifiphisher automates the process of de-authenticating a user from a legitimate internet access point, allows the user to connect to an ``evil twin'' access point that functions as a man-in-the-middle, and ultimately presents the target with a seemingly legitimate request that asks them to enter sensitive information.  We provide this information here specifically to argue the case that hacking of autonomous vehicles isn't just a possibility, but for all practical purposes, it is an inevitability.  As such, preventive strategies like those discussed in the main text will be essential for the safety of drivers.

\section*{An outlook on hacking autonomous vehicles}
Current estimates project approximately 150 million internet-connected vehicles on the road by 2020, with most sharing real-time data through cloud services\cite{ring2015}.  A comprehensive knowledge of vehicle localization, local traffic, weather, and infrastructure conditions would reduce congestion, lead to more efficient commutes, and potentially less accidents among many other possible benefits.  However, this increase in connectedness leads to a corresponding increase in vulnerability through mapping, entertainment, and productivity software.  As such, the National Highway Traffic Safety Administration lists 15 assessment criteria for significant software updates of vehicles and their related subsystems.  The first four criteria are (i) data recording and sharing, (ii) data privacy, (iii) system safety, and (iv) vehicle cybersecurity\cite{policy2016}, all of which are likely hacking targets.

A recent string of ``white hat hacks'' have demonstrated the existence of vulnerabilities in vehicle software through proof-of-concept demonstrations.  Chris Valasek and Charley Miller gave one of most well-known examples through a zero-day exploit of the entertainment system in a Jeep Cherokee to seize control of the target vehicle\cite{ring2015}.  In response, Fiat Chrysler Automobiles issued a software patch to 1.4 million affected vehicles.  In a significantly less-sophisticated demonstration, a low-cost ($\approx \$60$), low-power laser pulse generator was used to stop an autonomous vehicle by creating a ``ghost'' object in its drive path\cite{petit2015, harris2015}.  

Despite these concerning demonstrations, resources are being invested into the development of new software technology such as vehicle-to-vehicle (V2V) and vehicle-to-infrastructure (V2I) communications\cite{harding2014}, which allow the sharing of data through wireless networks.  These protocols will be essential for autonomous vehicles to maximize fuel efficiency, minimize traffic congestion, and prevent accidents.  However, this necessarily means vehicles will be communicating with one another, creating new opportunities for malicious hackers to exploit.  Indeed, thought leaders on the subject have expressed due concern.  Nevertheless, the U.S. DOT issued a proposed rule in December 2016 that would advance the deployment of connected vehicle technologies throughout the U.S. by mandating the deployment of V2V throughout the light vehicle fleet. 

Given the demonstrated hacks exploiting these new connected software systems, there is an urgent need for a better understanding of how collective motion can be weaponized.  As such, the anticipate-and-inoculate approach discussed in the main text will likely continue to progress alongside autonomous vehicle technologies, hopefully in a manner that proactively prevents any large-scale malicious hacks.

\section*{Equations of motion and numerics}
In the main text we provided a simplified form of the equations of motion where indices for vehicles were excluded.  We also combined the equations of motion for human- and autonomously-driven vehicles through the index $\alpha$, which takes the value $A$ for autonomously-driven vehicles and $H$ for human-driven vehicles.  Here, we explicitly write out the full equations of motion with indices and appropriate substitution for the repulsion force coefficient, as described in the main text.  Thus, for a simulation with $N_H$ human-driven vehicles and $N_A$ autonomously driven vehicles, we have
\begin{eqnarray}
\ddot{x}_i & = & F^{\rm propulsion}_i + F^{\rm repulsion}_i, \nonumber \\
F^{\rm propulsion}_i & = & \tau_{\alpha}^{-1} (v_{\alpha,i} - \dot{x}_i), \nonumber \\
F^{\rm repulsion}_i & = &  \left\{ 
\begin{array}{ll}
\left(\frac{v_{\alpha,i}}{\tau_{\alpha}} \right) \left[ \frac{ \left( 1 - \delta x_{i,i+1}/R \right)^{3/2} }{ ( 1- r/R )^{-3/2}} \right], & \delta x_{i,i+1} < R \\
0, & {\rm otherwise,}
\end{array}	\right. \nonumber \\
\label{eqSM1}
\end{eqnarray}
where the index $i = 1 \ldots N$ runs over all $N = N_A + N_H$ vehicles, the distance $\delta x_{i,i+1} = x_{i+1} - x_i$ is the separation between two consecutive cars within a lane, and the preferred speed $v_{\alpha,i}$ is drawn from a Gaussian distribution for the $N_H$ human-driven vehicles whereas autonomous vehicles are all assigned a single uniform value.  Simulations were performed on straight ``roads'' of length $L$ with periodic boundary conditions.  When a vehicle's position at time step $t$ was calculated to $x_i(t) = L + \delta L > L$, its position was stored as $x_i(t) = \delta L$.  As a result, the calculation of $\delta x_{i, i+1} = x(t)_{i+1} - x(t)_i$ takes into consideration the possibility that for vehicle $i$, the next vehicle ($i+1$) could be wrapped around back to the origin, and appropriate offsets are incorporated to avoid these edge effects.  For human-driven vehicles, the preferred velocities $v_{\alpha, i}$ were redrawn from the Gaussian distribution when they were wrapped around to the origin in order to keep faster vehicles from simply catching-up with the slowest vehicle and equilibrating into a single long chain of vehicles.  All simulations were performed in Python with a time step $\Delta t$ such that $\Delta t = (1/20) $ s, and a total of 10,000 time steps (500 s) were computed.

\begin{table*}
\caption{Empirical values for density, velocity, and flux.  The sources are: (1) Northbound 101 at Cabrillo Blvd., California, (2) Northbound 101 at 1$^{\rm st}$ St., California, (3) Southbound I85 exit 249C, GA, and (4) Southbound 101, Hollywood Freeway, California (DOT NGSIM dataset). Rows where the source is starred ($^*$) are used to generate the response time $\tau_H$ in Fig.~\ref{fig1}.  All entries are used in empirical measurements of $\Phi(\rho)$ in Fig.~\ref{fig2}.} \label{t1}
\begin{longtable*}{|c|ccc|ccc|ccc|}
\hline
 &  & $\rho$ &  &  & $v$ & & & $\Phi$ & \\
 &  & (cars/km/lane) &  &  & (km/h) & & & (cars/h/lane) & \\
\cline{2-10}
 &  & &  &  & & & & & \\
source & $1^{\rm st}$ quartile & median & $3^{\rm rd}$ quartile & $1^{\rm st}$ quartile & median & $3^{\rm rd}$ quartile & $1^{\rm st}$ quartile & median & $3^{\rm rd}$ quartile \\
\hline
1 & 10  & 13  & 16  & 71 & 76  & 83  & 719   & 1,004 & 1,285 \\ \hline
1 & 13  & 16  & 21  & 72 & 79  & 85  & 924   & 1,292 & 1,586 \\ \hline
1 & 29  & 32  & 37  & 35 & 37  & 39  & 1,019 & 1,209 & 1,364 \\ \hline
1 & 30  & 32  & 35  & 36 & 38  & 40  & 1,065 & 1,214 & 1,332 \\ \hline
2$^*$ & 85  & 95  & 117 & 5  & 6   & 9   & 509   & 660   & 873   \\ \hline
2 & 3   & 5   & 9   & 93 & 107 & 122 & 270   & 507   & 788   \\ \hline
3$^*$ & 104 & 127 & 149 & 3  & 6   & 8   & 432   & 693   & 899   \\ \hline
4$^*$ & 36 & 37 & 40 & 39  & 44   & 47   & 1,542   & 1,623   & 1,688   \\ \hline
\end{longtable*}
\end{table*}

\section*{Quantitative image analysis}
Daytime traffic videos in California (CA) were obtained from live traffic camera streams available on the Caltrans website\cite{livecam}.  Nighttime traffic video from Atlanta, Georgia (GA) was recorded manually from a conveniently located building near the highway off ramp using a Canon EOS Rebel DSLR.  Videos were analyzed using the OpenCV computer vision library and features of interest were detected through the Harris corner detection algorithm\cite{harris1988}.  Features were chosen and spatial averaging performed such that individual cars could be automatically localized and tracked within each video using the Lucas-Kanade\cite{lucas1981} algorithm.  Distances between lane markers were used to scale and correct for perspective distortions of the camera. Specifically, federal highway guidelines specify the use of 10 ft white markers with 40 ft spacing (center-to-center distance), though most drivers grossly underestimate this distance\cite{shaffer2008}. 

Empirical measurements [Fig.~\ref{fig2}] of the median and interquartile range for density $\rho$, velocity $v$, and flux $\Phi$ are provided here for reference [Table~\ref{t1}].

\section*{Calculation of human driver response time}
While the DOT data provided vehicle coordinates $x(t)$, we used quantitative image analysis on the first two video sources to obtain this information.  Differentiating and smoothing with a Savitzky-Golay filter allowed us to identify vehicles that stopped and accelerated over a time $t_f \approx 2 - 5$~s due to stop-and-go traffic.  To empirically determine the human-driver response time $\tau_H$, we assume that a driver in stopped traffic will accelerate when a sufficiently large gap between them and the next car becomes available.  The timescale during which the car adjusts its velocity then corresponds to the response timescale $\tau_H$. 

From this perspective, we define the moment a vehicle at rests starts moving as $t = t_0$.  From our image analysis and DOT data, we found that within a few seconds, most cars initially accelerate, but then experience a plateau in velocity.  At lower densities, this time to plateau is about 2 s, while for higher densities, it is about 5 s.  Defining this zero-to-plateau time interval as $t_f$, we extract $\tau_H$ by fitting measured vehicle position data to $\dot{x}(t-t_0)/\dot{x}(t_f-t_0) = 1 - e^{-t/\tau_H}$ with a non-linear least squares method.  In the main text, we simplified this expression by setting $t_0 = 0$ for convenience.  In higher-density traffic, we observed drivers tend to accelerate over longer periods of time, whereas in lower-density traffic, drivers tend to react faster.  From the empirical California (CA), Georgia (GA), and Department of Transportation (DOT) measurements of the human driver response time $\tau_H$ [Fig.~\ref{fig1}(c)], we calculated average $\tau_H  = 2.0$ s, which was used in simulations.

\section*{Validations of active matter model using one-lane traffic flow}
\begin{figure}
\includegraphics[scale=.5]{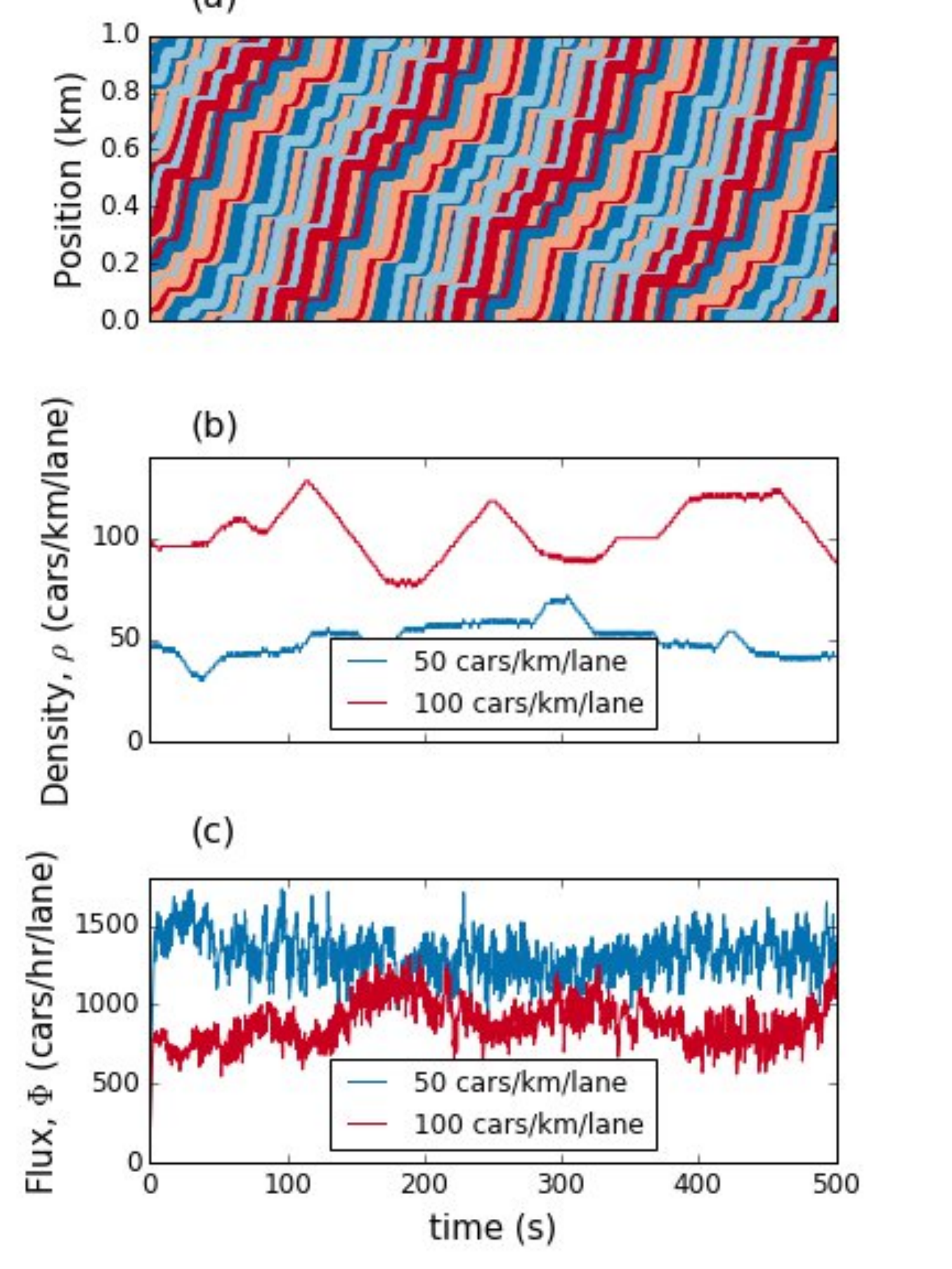}
\caption{Basic time-dependent quantification of simulated human-driven vehicles.  (a) Space-time diagram of vehicle trajectories.  Here, each randomly colored line (various shades of green) is a vehicle with average velocity $\Delta x / \Delta t > 0$.  When the local slope is flat, the vehicle has stopped due to the emergence of backwards-propagating density waves ($\Delta x / \Delta t < 0$, shaded orange).  Measurements of the (b) density and (c) flux show random statistical fluctuations that depend on the number of vehicles $N_H$ in the simulation.  Note, the total length of the simulated road $L = 1.0$ km, so that the average density $\approx N_H/L$. }
\label{figS1}
\end{figure}

Having defined the equations of motion for vehicles on a single lane road, we performed a series of test simulations to confirm the active matter model proposed here (e.g., Eq.~\ref{eq1}) reproduces basic phenomenology of vehicle traffic.  For example, simulating and plotting $x_i(t)$ for $N = N_H = 50$ vehicles immediately reveals the presence of backwards-propagating density waves, otherwise known as ``phantom traffic jams,'' that travel at velocity $\Delta x / \Delta t \approx -2$ m/s [Fig.~\ref{figS1}(a), dense horizontal bands].  Measurements of the time-dependent density $\rho(t)$ [Fig.~\ref{figS1}(b)] and flux $\Phi(t)$ [Fig.~\ref{figS1}(c)] provide further useful quantification of collective motion, as they are influenced by the balance between steady uniform flow and the emergence of phantom traffic jams.

\begin{figure*}
\includegraphics[scale=.98]{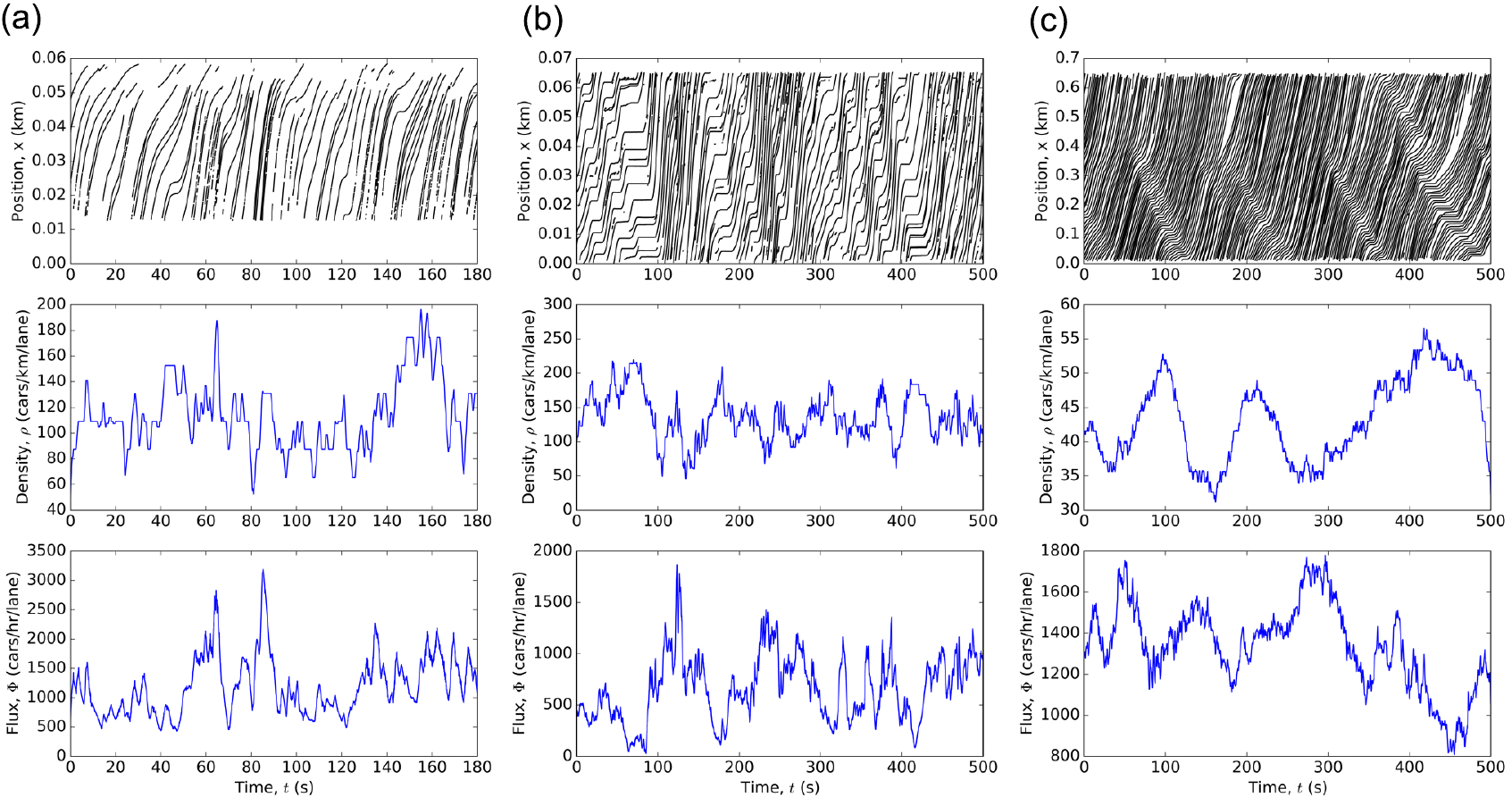}
\caption{Basic time-dependent quantification of empirical human-driven vehicles.  The three data sources in (a-c) correspond to the same three data sources described in the main text Fig.~\ref{fig1}.  The three rows of plots shown here mimic Fig.~\ref{figS1}, and show qualitative similarities.}
\label{figS2}
\end{figure*}

For comparison, we analyzed videos of single-lane traffic flow to generate the equivalent empirical observations.  We observed similar density waves with propagation speeds $\approx -3$ to $-6$ m/s, with qualitatively similar fluctuations in $\rho(t)$ and $\Phi(t)$ [Fig.~\ref{figS2}].  In the main text, we provide a more rigorous quantitative comparison between this empirical and simulation data [Fig.~\ref{fig2}].

\section*{MOBIL Lane changing}
To capture realistic highway traffic, we need a microscopic description to determine when vehicles should switch lanes in our active matter simulation.  While a variety of options have been established in the literature, here, we utilize the Minimizing Overall Breaking Induced By Lane changes (MOBIL) model\cite{treiber2006mobil,treiber2009modeling}.  This framework considers whether a vehicle and its neighbors would better match their preferred speed if the given vehicle changes lanes.  Defining the changes in acceleration between the next time step $t+1$ and the current time step $t$ due to a lane switch as
\begin{eqnarray}
\Delta \ddot{x}_i & = & \ddot{x}_i(t+1) - \ddot{x}_i(t), \nonumber \\
\Delta \ddot{x}_{i-1} & = & \ddot{x}_{i-1}(t+1) - \ddot{x}_{i-1}(t), \quad {\rm and} \nonumber \\
\Delta \ddot{x}_{j-1} & = & \ddot{x}_{j-1}(t+1) - \ddot{x}_{j-1}(t),
\end{eqnarray}
we can express the MOBIL condition as
\begin{equation}
\Delta \ddot{x}_i + p(\Delta \ddot{x}_{i-1} + \Delta \ddot{x}_{j-1}) > 0,
\end{equation}
where the subscript $i$ corresponds to the vehicle changing lanes, $i-1$ is the current vehicle behind the lane-changing vehicle at time $t$, and $j-1$ is the vehicle that will be behind the lane changing vehicle at $t+1$ if $i$ changes lanes.  The constant $p$ is referred to as the politeness factor.  A value $p = 1$ reflects the most considerate driver, where their decision to change lanes decreases the overall amount of braking among all three vehicles.  A value $p = 0$ reflects the most selfish driver, where a decision to change lanes allows for their own acceleration at the expense of other trailing vehicles.  A value $p > 1$ corresponds to the altruistic driver, where their decision to change lanes benefits trailing vehicles at the driver's expense.  And finally, a value $p < 0$ reflects a malicious driver, which deliberately forces trailing vehicles to use their brakes.

In our active matter model, we have self-propulsion and repulsive collision-avoidance forces, but only the repulsive force depend on the distance between vehicles. Thus, maximizing acceleration with MOBIL lane changing corresponds to minimizing the overall amount of repulsion in the system.  In all simulations, we choose $p = 1$, which corresponds to force minimization of a vehicle and its nearest neighbors.  Every time step, a vehicle has a 50\% chance to choose a neighboring lane to change into.  This is done to avoid artificial states, where all vehicles simultaneously switch into the same lane, leaving large empty gaps.  The left-most ($\ell = 1$) and right-most ($\ell = 3$) lanes, can change into the center ($\ell = 2$) lane, whereas the center lane can choose between the other two lanes with equal probability.  Within fluctuations, we find that different $p$ values do not affect macroscopic flux-densities measurements.  Likewise, in the presence of disabled vehicles, $p$ does not change the underlying percolation-geometric transition discussed in the main text.

\section*{Nagel-Schrekenberg cellular automata models}
To confirm the model-independence of our results, we additionally perform Nagel-Schrekenberg\cite{nagel1992} cellular automata simulations.  These correspond to simulations of vehicles on a lattice of fixed cell length $d_s$, where vehicles are updated according to certain rules that mimic realistic driving. The basic rules are (i) vehicles accelerate to a maximum velocity, (ii) unless there is a vehicle in front of them, (iii) in which case the vehicle slows down accordingly, and finally, (iv) vehicles are slowed down at random, to incorporate driving variances. Here, we performed simulations corresponding to two different cell lengths, 4.5 and 7.0~m respectively.  The maximum velocity of vehicles are in units of $d_s$ m/s, where $d_s$ is the cell length.  For a 1,000~m road, these distances correspond to $\approx 143$ and 222 sites.  Additionally, we set the maximum velocity $v_0$ to 7 units per time step, which translates to 29~m/s (65 mi/h). 
\begin{figure}
\includegraphics[scale=.6]{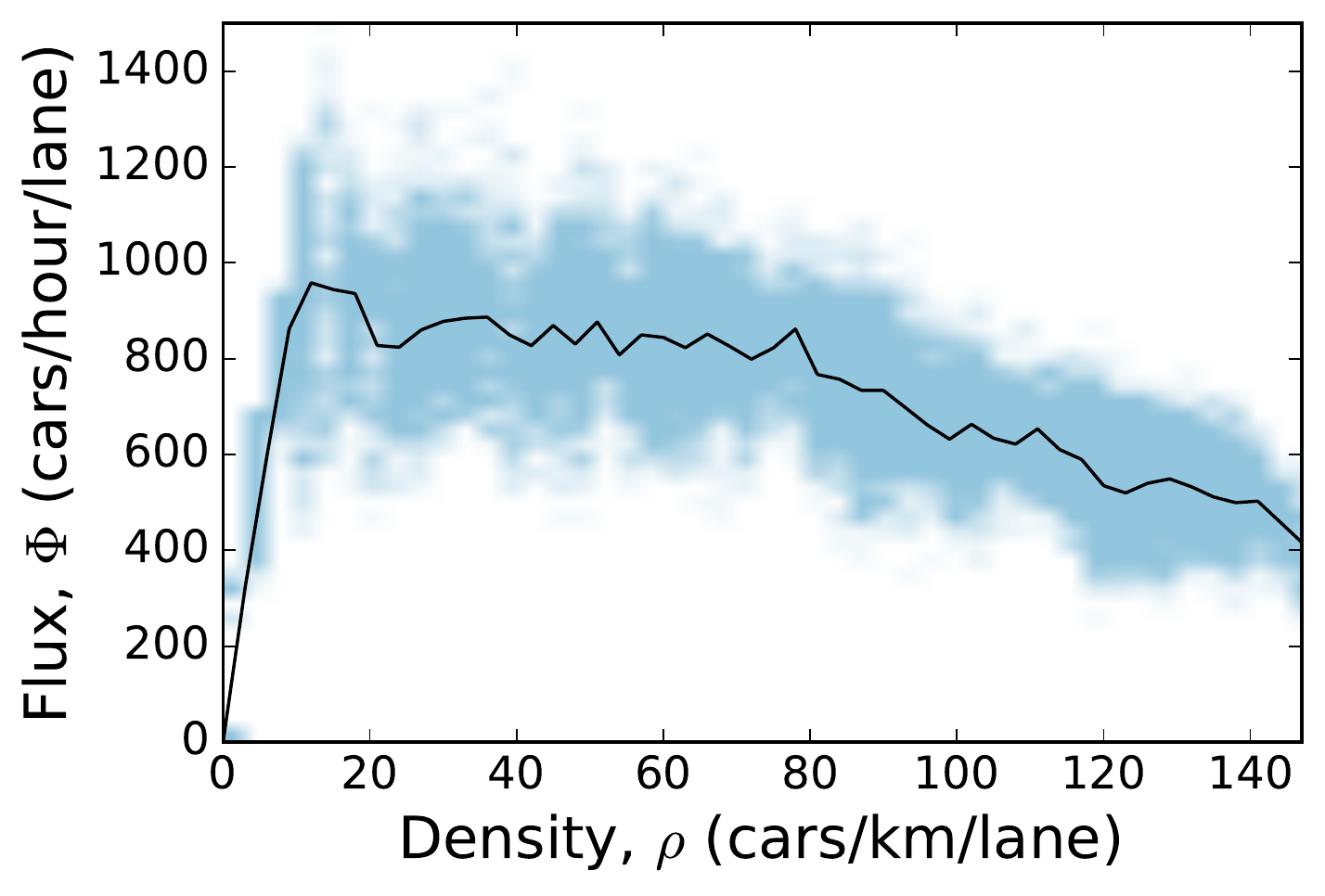}
\caption{Flux-density curve for Nagel-Schrekenberg cellular automata simulation with cell length equal to 4.5~m.  Black line denotes average value of flux while blue shaded band indicates fluctuations about this mean.}
\label{figS4}
\end{figure}

At every time step, a random number $p \in [0,1]$ is computed, and if $p > 0.5$, the following rules apply.  The update rule for vehicle $\alpha$ is $v_\alpha(t + 1)=\textrm{min}(v_{\alpha}+ 1,v_0,s_\alpha)$, where $s_\alpha$ is the distance from car $\alpha$ to the car in front. This update rule ensures that at every step the velocity is increased by 1, until the maximum velocity $v_0$ is reached, unless obstructed by a leading vehicle, in which case speed is reduced accordingly.  If $p < 0.5$, the vehicle decreases its velocity by 1, without going through the above update rule - this is the critical component that gives rise to driving variances and traffic jams in this model. The flux-density curve for 4.5~m cell length simulations is shown in Fig.~\ref{figS4}. While the flux-density curve has clear differences with continuous laned active matter simulations [Fig.\ref{fig2}b], percolation probabilities fall right on the appropriate discrete geometrical predictions [Fig.\ref{fig4}(a)].  This validates our claim that percolation results are model independent.

\onecolumngrid


\begin{thebibliography}{10}

\bibitem{litman2014}
Todd Litman.
\newblock Autonomous vehicle implementation predictions.
\newblock {\em Victoria Transport Policy Institute}, 28, 2014.

\bibitem{Greenberg2015Hackers}
Andy Greenberg.
\newblock Hackers remotely kill a jeep on the highway -- with me in it, 2015.
\newblock
  \url{https://www.wired.com/2015/07/hackers-remotely-kill-jeep-highway}
  [Online; accessed 2-August-2017].

\bibitem{Greenberg2016The}
Andy Greenberg.
\newblock The jeep hackers are back to prove car hacking can get much worse,
  2016.
\newblock
  \url{https://www.wired.com/2016/08/jeep-hackers-return-high-speed-steering-acceleration-hacks}
  [Online; accessed 2-August-2017].

\bibitem{parkinson2017cyber}
Simon Parkinson, Paul Ward, Kyle Wilson, and Jonathan Miller.
\newblock Cyber threats facing autonomous and connected vehicles: future
  challenges.
\newblock {\em IEEE Transactions on Intelligent Transportation Systems},
  18(11):2898--2915, 2017.

\bibitem{amoozadeh2015security}
Mani Amoozadeh, Arun Raghuramu, Chen-Nee Chuah, Dipak Ghosal, H~Michael Zhang,
  Jeff Rowe, and Karl Levitt.
\newblock Security vulnerabilities of connected vehicle streams and their
  impact on cooperative driving.
\newblock {\em IEEE Communications Magazine}, 53(6):126--132, 2015.

\bibitem{ramaswamy2010mechanics}
Sriram Ramaswamy.
\newblock The mechanics and statistics of active matter.
\newblock {\em The Annual Review of Condensed Matter Physics}, 1:323--45, 2010.

\bibitem{marchetti2013hydrodynamics}
M~Cristina Marchetti, JF~Joanny, S~Ramaswamy, TB~Liverpool, J~Prost, Madan Rao,
  and R~Aditi Simha.
\newblock Hydrodynamics of soft active matter.
\newblock {\em Reviews of Modern Physics}, 85(3):1143, 2013.

\bibitem{thomas2015fraction}
CC~Thomas and Douglas~J Durian.
\newblock Fraction of clogging configurations sampled by granular hopper flow.
\newblock {\em Physical review letters}, 114(17):178001, 2015.

\bibitem{zuriguel2014clogging}
Iker Zuriguel, Daniel~Ricardo Parisi, Ra{\'u}l~Cruz Hidalgo, Celia Lozano,
  Alvaro Janda, Paula~Alejandra Gago, Juan~Pablo Peralta, Luis~Miguel Ferrer,
  Luis~Ariel Pugnaloni, Eric Cl{\'e}ment, et~al.
\newblock Clogging transition of many-particle systems flowing through
  bottlenecks.
\newblock {\em Scientific reports}, 4:7324, 2014.

\bibitem{helbing2001}
Dirk Helbing.
\newblock Traffic and related self-driven many-particle systems.
\newblock {\em Reviews of Modern Physics}, 73:1067--1141, Dec 2001.

\bibitem{treiber2013}
Martin Treiber and Arne Kesting.
\newblock Traffic flow dynamics.
\newblock {\em Traffic Flow Dynamics: Data, Models and Simulation,
  Springer-Verlag Berlin Heidelberg}, 2013.

\bibitem{lwr}
M.~J. Lighthill and G.~B. Whitham.
\newblock On kinematic waves. ii. a theory of traffic flow on long crowded
  roads.
\newblock {\em Proceedings of the Royal Society of London A: Mathematical,
  Physical and Engineering Sciences}, 229(1178):317--345, 1955.

\bibitem{newell2002}
Gordon~Frank Newell.
\newblock A simplified car-following theory: a lower order model.
\newblock {\em Transportation Research Part B: Methodological}, 36(3):195--205,
  2002.

\bibitem{treiber2000}
Martin Treiber, Ansgar Hennecke, and Dirk Helbing.
\newblock Congested traffic states in empirical observations and microscopic
  simulations.
\newblock {\em Physical Review E}, 62(2):1805, 2000.

\bibitem{nagel1992}
Kai Nagel and Michael Schreckenberg.
\newblock A cellular automaton model for freeway traffic.
\newblock {\em Journal de Physique I}, 2(12):2221--2229, 1992.

\bibitem{nagatani2002physics}
Takashi Nagatani.
\newblock The physics of traffic jams.
\newblock {\em Reports on Progress in Physics}, 65(9):1331, 2002.

\bibitem{ovm-bando1994}
M~Bando.
\newblock M. bando, k. hasebe, a. nakayama, a. shibata, and y. sugiyama, jpn.
  j. ind. appl. math. 11, 203 (1994).
\newblock {\em Jpn. J. Ind. Appl. Math.}, 11:203, 1994.

\bibitem{silverberg2013collective}
Jesse~L Silverberg, Matthew Bierbaum, James~P Sethna, and Itai Cohen.
\newblock Collective motion of humans in mosh and circle pits at heavy metal
  concerts.
\newblock {\em Physical Review Letters}, 110(22):228701, 2013.

\bibitem{bottinelli2016emergent}
Arianna Bottinelli, David~TJ Sumpter, and Jesse~L Silverberg.
\newblock Emergent structural mechanisms for high-density collective motion
  inspired by human crowds.
\newblock {\em Physical Review Letters}, 117(22):228301, 2016.

\bibitem{bottinelli2017using}
Arianna Bottinelli and Jesse~L Silverberg.
\newblock How to: Using mode analysis to quantify, analyze, and interpret the
  mechanisms of high-density collective motion.
\newblock {\em Frontiers in Applied Mathematics and Statistics}, 3:26, 2017.

\bibitem{landau1959course}
Lev~Davidovich Landau and Eugin~M Lifshitz.
\newblock {\em Course of Theoretical Physics Vol 7: Theory and Elasticity}.
\newblock Pergamon Press, 1959.

\bibitem{livecam}
\url{http://www.dot.ca.gov/video/index.html}.

\bibitem{ngsim2005}
United States~Department of~Transportation Federal High~Administration.
\newblock Ngsim datasets, 2005.
\newblock \url{https://www.its-rde.net/index.php/rdedataenvironment/10023}
  [Online; accessed 2-August-2017].

\bibitem{alexiadis2007model}
Vassili Alexiadis, James Colyar, and John Halkias.
\newblock A model endeavor.
\newblock {\em Public Roads}, 70(4), 2007.

\bibitem{montanino2013making}
Marcello Montanino and Vincenzo Punzo.
\newblock Making ngsim data usable for studies on traffic flow theory:
  Multistep method for vehicle trajectory reconstruction.
\newblock {\em Transportation Research Record: Journal of the Transportation
  Research Board}, 2390:99--111, 2013.

\bibitem{treiber2006mobil}
Martin Treiber and Dirk Helbing.
\newblock Mobil: General lane-changing model for car-following models.
\newblock {\em Dispon{\i}vel em http://www. mtreiber.
  de/publications/MOBIL\_TRB. pdf, Acesso em dezembro de}, 2009, 2016.

\bibitem{treiber2009modeling}
Martin Treiber and Arne Kesting.
\newblock Modeling lane-changing decisions with mobil.
\newblock {\em Traffic and Granular Flow’07}, pages 211--221, 2009.

\bibitem{kirkpatrick1973percolation}
Scott Kirkpatrick.
\newblock Percolation and conduction.
\newblock {\em Reviews of modern physics}, 45(4):574, 1973.

\bibitem{aharony2003introduction}
Amnon Aharony and Dietrich Stauffer.
\newblock {\em Introduction to percolation theory}.
\newblock Taylor \& Francis, 2003.

\bibitem{bollobas2006percolation}
B{\'e}la Bollob{\'a}s and Oliver Riordan.
\newblock {\em Percolation}.
\newblock Cambridge University Press, 2006.

\bibitem{silverberg2009model}
Jesse~L Silverberg.
\newblock A model for conductive percolation in ordered nanowire arrays.
\newblock {\em Journal of Applied Physics}, 105(4):044306, 2009.

\bibitem{cejas2017particle}
Cesare~Mikhail Cejas, Fabrice Monti, Marine Truchet, Jean-Pierre Burnouf, and
  Patrick Tabeling.
\newblock Particle deposition kinetics of colloidal suspensions in
  microchannels at high ionic strength.
\newblock {\em Langmuir}, 2017.

\bibitem{recode2017auto}
Nabeel Hyatt.
\newblock Autonomous driving is here, and it's going to change everything,
  2017.
\newblock
  \url{https://www.recode.net/2017/4/19/15364608/autonomous-self-driving-cars-impact-disruption-society-mobility}
  [Online; accessed 2-August-2017].

\bibitem{cnbc2017self}
Joel Barbier.
\newblock Self-driving cars will disrupt more than the auto industry. here are
  the winners and losers, 2017.
\newblock
  \url{https://www.cnbc.com/2017/05/03/self-driving-cars-will-disrupt-10-industries-commentary.html}
  [Online; accessed 2-August-2017].

\bibitem{wapo2016how}
Brian Fung.
\newblock How driverless cars could kill the speeding ticket - and rob your
  city, 2016.
\newblock
  \url{https://www.washingtonpost.com/news/the-switch/wp/2016/01/22/how-driverless-cars-could-kill-the-speeding-ticket-and-rob-your-city}
  [Online; accessed 2-August-2017].

\bibitem{center2016auto}
University of~Michigan Center~for Sustainable~Systems.
\newblock Autonomous vehicles factsheet, 2016.
\newblock \url{http://css.umich.edu/factsheets/autonomous-vehicles-factsheet}
  [Pub. No. CSS16-18; Online; accessed 2-August-2017].

\bibitem{npr2015map}
Quoctrung Bui.
\newblock Map: The most common job in every state, 2015.
\newblock
  \url{http://www.npr.org/sections/money/2015/02/05/382664837/map-the-most-common-job-in-every-state}
  [Online; accessed 2-August-2017].

\bibitem{Greenberg2017How}
Andy Greenberg.
\newblock How an entire national became russia's test lab for cyberwar, 2017.
\newblock \url{https://www.wired.com/story/russian-hackers-attack-ukraine}
  [Online; accessed 2-August-2017].

\bibitem{gjelten2010cyberworm}
Tom Gjelten.
\newblock Cyberworm's origins unclear, but potential is not, 2010.
\newblock \url{http://www.npr.org/templates/story/story.php?storyId=130162219}
  [Online; accessed 2-August-2017].

\bibitem{shapiro2016All}
Ari Shapiro.
\newblock Documentary explores the cyber-war secrets of stuxnet, 2016.
\newblock
  \url{http://www.npr.org/2016/07/04/484713086/documentary-explores-the-cyber-war-secrets-of-stuxnet}
  [Online; accessed 2-August-2017].

\bibitem{chappell2017ransomware}
Bill Chappell and Maggie Penman.
\newblock Ransomware attacks ravage computer networks in dozens of countries,
  2017.
\newblock
  \url{http://www.npr.org/sections/thetwo-way/2017/05/12/528119808/large-cyber-attack-hits-englands-nhs-hospital-system-ransoms-demanded}
  [Online; accessed 2-August-2017].

\bibitem{ring2015}
Tim Ring.
\newblock Connected cars--the next target for hackers.
\newblock {\em Network Security}, 2015(11):11--16, 2015.

\bibitem{policy2016}
Federal Automated~Vehicles Policy.
\newblock Accelerating the next revolution in roadway safety, nhtsa, us dept.
\newblock {\em Transportation}, 2016.

\bibitem{petit2015}
Jonathan Petit and Steven~E Shladover.
\newblock Potential cyberattacks on automated vehicles.
\newblock {\em IEEE Transactions on Intelligent Transportation Systems},
  16(2):546--556, 2015.

\bibitem{harris2015}
Mark Harris.
\newblock Researcher hacks self-driving car sensors.
\newblock {\em IEEE Spectrum}, 9, 2015.

\bibitem{harding2014}
John Harding, Gregory Powell, Rebecca Yoon, Joshua Fikentscher, Charlene Doyle,
  Dana Sade, Mike Lukuc, Jim Simons, and Jing Wang.
\newblock Vehicle-to-vehicle communications: Readiness of v2v technology for
  application.
\newblock Technical report, National Highway Traffic Safety Administration,
  2014.

\bibitem{harris1988}
Chris Harris and Mike Stephens.
\newblock A combined corner and edge detector.
\newblock In {\em Alvey Vision Conference}, volume~15, pages 10--5244.
  Manchester, UK, 1988.

\bibitem{lucas1981}
Bruce~D Lucas, Takeo Kanade, et~al.
\newblock An iterative image registration technique with an application to
  stereo vision.
\newblock {\em Proceedings DARPA Image Understanding Workshop}, 1981.

\bibitem{shaffer2008}
Dennis~M. Shaffer, Andrew~B. Maynor, and Windy~L. Roy.
\newblock The visual perception of lines on the road.
\newblock {\em Perception {\&} Psychophysics}, 70(8):1571--1580, Nov 2008.

\end{thebibliography}
\end{document}